\documentclass[amsmath,preprint]{revtex4}
\usepackage{graphicx}

\newcommand{\omb}{\omega_\text{b}}
\renewcommand{\vec}[1]{\boldsymbol{#1}}

\begin{document}

\title{Phase-space geometry of the generalized Langevin equation}
\author{Thomas Bartsch}
\affiliation{Department of Mathematical Sciences, Loughborough University, Loughborough LE11 3TU, UK}

\begin{abstract}
The generalized Langevin equation is widely used to model the influence of a heat bath upon a reactive system. This equation will here be studied from a geometric point of view. A dynamical phase space that represents all possible states of the system will be constructed, the generalized Langevin equation will be formally rewritten as a pair of coupled ordinary differential equations, and the fundamental geometric structures in phase space will be described. It will be shown that the phase space itself and its geometric structure depend critically on the preparation of the system: A system that is assumed to have been in existence for ever has a larger phase space with a simpler structure than a system that is prepared at a finite time. These differences persist even in the long-time limit, where one might expect the details of preparation to become irrelevant.
\end{abstract}
\maketitle

\section{Introduction}

The influence of a solvent or another complex environment on a chemical motion has routinely been modeled using the Langevin equation or its generalization\cite{Kramers40,Zwanzig61,Zwanzig61a,Prigogine61,Mori65,Fox78,Haenggi90,Zwanzig01}. In these equations, complicated many-body effects are approximated by three modifications of the equations of motion that describe the intrinsic dynamics of the reactive system: (i) a change of the potential energy surface that governs the dynamics, (ii) a stochastic force, and (iii) a dissipative friction force. The latter two forces are related by fluctuation-dissipation theorems. 

The simplest assumption one can make about the noise is that it is white noise, i.e., the strengths of the noise force at different times are statistically independent. This assumption requires that the dynamics of the heat bath, which determines the correlation time of the fluctuating force, takes place on much shorter time scales than the dynamics of the chosen system coordinates, viz., the reaction coordinate and any other solute or solvent coordinate coupled to it. In chemical applications, however, the dynamical time scales of the system and the bath are usually comparable, so that it is often essential to allow temporal correlations of the noise. Fluctuation-dissipation theorems \cite{Zwanzig61a,Haenggi90} then require that the form of the dissipation term must be adapted in such a way that the strength of the damping force depends not only on the instantaneous velocity, but on the entire history of the trajectory. Thus, the inclusion of correlated noise forces a fundamental change in  the mathematical structure of the theory: Whereas the Langevin equation with white noise is a stochastic differential equation, the generalized Langevin equation that models correlated noise is an integro-differential equation that is considerably harder to treat.

One of the most fundamental properties of a dynamical system is its phase space, which represents all possible states of the system in such a way that every phase space point corresponds to one and only one state. Recent developments have highlighted the importance of adopting a phase space view in dynamical studies of chemical reactivity that are based on Transition State Theory (TST): Suitable phase space structures represent the ideal recrossing-free dividing surface between reactants and products that was long sought for in TST \cite{Pechukas77a,Komatsuzaki96,Komatsuzaki97,Komatsuzaki00,Jaffe99,Wiggins01,Uzer02}. In addition to  the dividing surface itself, the phase space view of TST provides surfaces in phase space that separate reactive from non-reactive trajectories and that can be used in an efficient calculation of reaction rates \cite{Ozorio90,DeLeon91a,DeLeon91b,Uzer02,Waalkens04c,Waalkens05a,Waalkens05b}. Although all these structures are fundamentally classical objects, they retain their importance for semiclassical treatments of the reaction dynamics \cite{Hernandez93,Hernandez94}.

A recent series of papers \cite{Bartsch05b,Bartsch05c,Bartsch06a,Kawai07a,Bartsch08,Bartsch08a} generalized the phase space view of TST to reactions that take place under the influence of time-dependent driving forces or external noise. It was shown that all important phase space structures of TST remain intact, but move through phase space stochastically. For deterministically driven systems and systems under the influence of white noise that are described by the Langevin equation, this work could rely on an explicit representation of the phase space: The phase space consists of all possible values of coordinates and momenta (or velocities). In these cases, the investigation can immediately turn to the problem of describing the dynamical and geometric structures in phase space that encode the important features of the dynamics. The generalized Langevin equation, however, has a more complicated mathematical structure, and it is not immediately obvious what the phase space is or even how many dimensions it has. Several constructions of appropriate phase spaces have been suggested in the literature, e.g. in \cite{Dygas86,Straub86,Frishman93,Martens02}. It will here be shown that the phase space depends critically on the way in which the system is prepared. Realistically, every system has come into existence at a certain time in the past, but if that time is sufficiently remote, one may simplify the description by assuming that the system has existed for ever. Perhaps surprisingly, this assumption changes the phase space drastically. A finite time of preparation, no matter how remote, imposes restrictions that reduce the dimension of phase space and make its structure more complex. A detailed description of how assumptions about preparation influence phase space structure even in the long term limit, when one might expect them to be forgotten, is the main aim of this paper.

The generalized Langevin equation is usually written in the form
\begin{equation}
\label{GLE_finite_inh}
\ddot {\vec q} = -\nabla V(\vec q) - \int_{0}^t \gamma(t-s)\,\dot {\vec q}(s)\, ds + \vec\xi_\alpha(t),
\end{equation}
where $V(\vec q)$ is a potential energy surface that governs the dynamics of the process under study, $\vec\xi_\alpha(t)$ is an external ``noise'' force exerted by the environment of which typically only the statistical properties are known. The integral represents a friction force that the system experiences due to its interaction with the environment. The strength of this damping force at any given time depends on the prehistory of the system from time $t=0$ in a manner that is described by the friction kernel $\gamma(t)$.  In thermal equilibrium, the friction kernel is related to the correlation function of the noise by a fluctuation-dissipation theorem~\cite{Zwanzig61a,Haenggi90}. The form of the friction term used in Eq.~\eqref{GLE_finite_inh} assumes that the system is prepared ``from scratch'' at time $t=0$ and that the history of the system before that time does not have any influence on its future development. That assumption, however, is not always appropriate. In some contexts (see, e.g., \cite{Bartsch05c}, where it was convenient to impose a boundary condition at $t\to-\infty$) one might want to assume that the system has been in existence for a long time, and thus to replace the GLE~\eqref{GLE_finite_inh} by
\begin{equation}
\label{GLE_inf_inh}
\ddot {\vec q} = -\nabla V(\vec q) - \int_{-\infty}^t \gamma(t-s)\,\dot {\vec q}(s)\, ds + \vec \xi_\alpha(t).
\end{equation}
This change might at first glance appear trivial. In particular, one might expect that for sufficiently large time $t$ the two versions of the GLE become indistinguishable because the difference between the two friction terms will tend to zero. That this conclusion is false becomes immediately clear when one considers the number of initial or boundary conditions that can be imposed on the solutions of the two GLE: Eq.~\eqref{GLE_finite_inh} poses an initial-value problem. Once the position and velocity at time $t=0$ are prescribed, the future of the trajectory is completely determined. To determine a trajectory of Eq.~\eqref{GLE_inf_inh}, by contrast, one has to prescribe its entire history from $t=-\infty$ to the present. The prescribed history of the trajectory is constrained by the requirement that it must satisfy the GLE. The number of free parameters one has to choose can be finite or infinite, but it is always larger than two. There is therefore a qualitative difference in dynamics between the two versions of the GLE that is manifest in the dimension of their phase spaces and that does not disappear over time.

It is the purpose of the present paper to elucidate the difference in phase space structure that is brought about by the seemingly innocuous change in the structure of the damping term. The development will be built on a representation of phase space for the GLE that was proposed by Martens~\cite{Martens02} and that applies to the infinite-time GLE~\eqref{GLE_inf_inh}. A phase space for the finite-time GLE~\eqref{GLE_finite_inh} will be obtained as a time-dependent two-dimensional surface in the phase space of the infinite-time GLE, and the dynamics in that surface are given are described by time-dependent ordinary differential equations. In this way, it will be shown that the dynamics of the infinite-time GLE are simpler than those of its more common finite-time counterpart.

In order to analyze the phase space structures of the GLE, we will for the sake of simplicity assume that the Brownian particle is moving in a one-dimensional configuration space, i.e., the position vector $\vec q$ simplifies to a scalar coordinate $q$. The formalism to be developed generalizes to higher-dimensional problems in a straightforward manner. We will also assume that the motion takes place in the vicinity of a parabolic barrier and that therefore $-\nabla V(q)=\omb^2 q$. This approximation, which covers the case that is most important in reaction rate theory, allows us to solve the equations of motion explicitly and to check in this way that the phase space that is to be constructed has the correct dimension. Higher-order corrections to the potential will deform the phase space structures, but leave them qualitatively unchanged, as is discussed in more detail at the end of Section~\ref{sec:noise}. Throughout most of this paper, we will also leave out the noise terms in Eq.~\eqref{GLE_finite_inh} and~\eqref{GLE_inf_inh} and study the autonomous equations
\begin{equation}
\label{GLE_finite}
\ddot q = \omb^2 q - \int_{0}^t \gamma(t-s)\,\dot q(s)\, ds 
\end{equation}
which will henceforth be called the finite-time GLE, and
\begin{equation}
\label{GLE_inf}
\ddot q = \omb^2 q - \int_{-\infty}^t \gamma(t-s)\,\dot q(s)\, ds ,
\end{equation}
which will be referred to as the infinite-time GLE. This omission is justified because we are mainly interested in a qualitative theory of the phase space structure. The modifications that are required when the noise is included will be described in Section~\ref{sec:noise}.

The structure of this paper is as follows: Section~\ref{sec:solInf} describes the general solution of  the linearized infinite-time GLE, and thus implicitly its phase space. Section~\ref{sec:extPhase} reviews the explicit construction of a phase space for the GLE that was given in~\cite{Martens02} and that describes the dynamics of the infinite-time GLE. The general solution of the finite-time GLE is derived in Section~\ref{sec:solFin}, where it is also shown how the finite-time damping term restricts the number of solutions as compared to its infinite-time counterpart. In Section~\ref{sec:finitePhase}, a phase space for the finite-time GLE is constructed as a time-dependent subspace of the phase space for the infinite-time GLE. Section~\ref{sec:exp} illustrates the construction in the simple case of an exponentially decaying friction kernel, and it explains why the results obtained in this special case are typical of more general friction kernels. Finally, in Section~\ref{sec:noise} the noise terms is reintroduced into the GLE, and the influence of the noise on the phase space structure is described.

\section{Solution of the infinite-time GLE}
\label{sec:solInf}

It is well known~(see, e.g., \cite{Zwanzig01}) that the finite-time GLE~\eqref{GLE_finite} can be solved by a one-sided Laplace transform, the infinite-time GLE~\eqref{GLE_inf} by a two-sided Laplace transform. However, the two-sided Laplace transform is only defined for functions that decay sufficiently fast in both the past and the future. In particular, the two-sided Laplace transform of an exponential function, which is expected as a solution of the GLE, is undefined. Thus, although the calculation formally leads to the correct result, it seems preferable to avoid using Laplace transforms. We will here work directly with the GLE. The calculation will here be presented in detail in order to highlight the differences between the finite-time and the infinite-time GLE.

If we substitute the ansatz $q(t)=e^{\epsilon t}$ into the GLE~\eqref{GLE_inf}, the damping term takes the form
\begin{align}
\label{damp_inf}
\int_{-\infty}^t \gamma(t-s) \,\dot q(s) \,ds &=
	\epsilon \int_{-\infty}^t \gamma(t-s)\, e^{\epsilon s}\,ds \nonumber\\
	&= \epsilon e^{\epsilon t} \int_0^\infty \gamma(r) \,e^{-\epsilon r}\, dr \nonumber \\
	&= \epsilon e^{\epsilon t} \hat\gamma(\epsilon) ,
\end{align}
where $\hat\gamma(\epsilon)$ is the Laplace transform of the friction kernel $\gamma(t)$, which is defined by the integral in~\eqref{damp_inf}. The GLE goes over into the nonlinear eigenvalue equation
\begin{equation}
	\label{evalEq}
	\epsilon^2 + \epsilon \hat\gamma(\epsilon) - \omb^2 = 0
\end{equation}
which determines the values of $\epsilon$ for which the exponential $q(t)=e^{\epsilon t}$ is a solution of the GLE. Because the GLE~\eqref{GLE_inf} describes the dynamics in the vicinity of a potential barrier, we will in general find one positive real eigenvalue that corresponds to an unstable mode: The system slides down the barrier. This eigenvalue is the well-known Grote-Hynes reaction frequency \cite{Grote80} that determines the reaction rate in the space-diffusion limited regime. The unstable mode can be interpreted as a collective reaction coordinate that describes correlated motion of the reactive subsystem and the bath. It can be constructed either from an explicit model of the heat bath \cite{Pollak86,Pollak89} or from a continuum model \cite{Graham90}. In addition to the unstable mode, there will be several eigenvalues with negative real parts that describe stable bath modes. Because the eigenvalue equation~\eqref{evalEq} is analytic and is real on the real axis, complex eigenvalues must occur in complex conjugate pairs.

The general solution of the infinite-time GLE~\eqref{GLE_inf} can be written as a linear superposition
\begin{equation}
	\label{gensol_inf}
	q(t) = \sum_j c_j e^{\epsilon_j t},
\end{equation}
where the eigenvalues $\epsilon_j$ are the solutions of the eigenvalue equation~\eqref{evalEq} and $c_j$ are arbitrary constants. They are subject to the condition that $q(t)$ should be real, i.e., if there are complex conjugate eigenvalues, the corresponding prefactors must also be complex conjugate. Consequently, the general solution~\eqref{gensol_inf} has as many free real parameters as there are solutions to the eigenvalue equation. This number is the phase space dimension of the infinite-time GLE.

The previous discussion describes the general solution of the GLE~\eqref{GLE_inf} correctly if all eigenvalues are simple. If the eigenvalue equation~\eqref{evalEq} has multiple solutions, one might expect that additional solutions should exist. Indeed, it can easily be checked that the ansatz
\begin{equation}
	\label{multAnsatz}
	q(t) = t^k e^{\epsilon t}
\end{equation}
with some positive integer~$k$ provides a solution of the infinite-time GLE if $\epsilon$ is an $n$th order zero of the eigenvalue equation~\eqref{evalEq} and $k<n$. In this way, an $n$-fold eigenvalue~$\epsilon$ gives rise to the $n$ solutions
\begin{equation*}
	e^{\epsilon t}, \quad t e^{\epsilon t}, \quad \dots, \quad t^{n-1} e^{\epsilon t}.
\end{equation*}
The total number of independent solutions of the infinite-time GLE~\eqref{GLE_inf} is therefore the number of solutions to the eigenvalue equation~\eqref{evalEq}, counted with multiplicities. Thus, this number remains unchanged if parameters in the GLE such as the damping strength are varied continuously in such a way that eigenvalues coincide.

\section{The extended phase space}
\label{sec:extPhase}

Martens~\cite{Martens02}, following earlier approaches~\cite{Dygas86,Straub86,Frishman93}, suggested a construction of the dynamical phase space under the assumption that the friction kernel~$\gamma(t)$ satisfies a linear differential equation with constant coefficients of the form
\begin{equation}
	\label{gammaEq}
	\gamma^{(n+1)}(t) + \sum_{i=0}^n a_i \gamma^{(i)}(t) = 0,
\end{equation}
where
\begin{equation*}
	\gamma^{(i)}(t) = \frac{d^i \gamma(t)}{dt^i}
\end{equation*}
and $a_0,\dots,a_n$ are arbitrary real constants. If the friction kernel satisfies this condition, one can obtain an explicit representation of the dynamical phase space by introducing the customary momentum coordinate $p=\dot q$ and the $n+1$ auxiliary variables
\begin{equation}
	\label{zetaDef}
	\begin{split}
		\zeta_0(t) &= \int_{-\infty}^t \gamma(t-s) \,p(s) \,ds, \\
		\zeta_1(t) &= \int_{-\infty}^t \dot\gamma(t-s) \,p(s) \,ds, \\
		&\dots \\
		\zeta_n(t) &= \int_{-\infty}^t \gamma^{(n)}(t-s) \, p(s) \, ds.
	\end{split}
\end{equation}
These definitions differ from those proposed in~\cite{Martens02} in that the lower limit of integration was set to $-\infty$ in order to describe the dynamics of the infinite-time GLE.

With the help of the variables~$\zeta_i$ the GLE can be rewritten as a system of linear ordinary differential equations
\begin{equation}
	\label{odes}
	\begin{split}
		\dot q &= p, \\
		\dot p &= \omb^2 q-\zeta_0, \\
		\dot \zeta_i &= \gamma^{(i)}(0)\,p + \zeta_{i+1} \qquad \text{for }0\le i < n,\\
		\dot \zeta_n &= \gamma^{(n)}(0)\,p - \sum_{i=0}^n a_i \zeta_i.
	\end{split}
\end{equation}
In the last of these equations, the differential equation~\eqref{gammaEq} of the friction kernel has been used. If these coordinates are gathered into the vector
\begin{equation}
	\vec z = \begin{pmatrix}
			q \\ p \\ \zeta_0 \\ \vdots \\ \zeta_n
			\end{pmatrix},
\end{equation}
the equation system~\eqref{odes} takes the form
\begin{equation}
	\label{odes_vec}
	\dot{\vec z} = M\,\vec z
\end{equation}
with the matrix
\begin{equation}
	\label{Mdef}
	M = \begin{pmatrix}
			0 & 1 & 0 & 0 & 0 & \dots & 0\\
			\omb^2 & 0 & -1 & 0 & 0 & \dots & 0\\
			0 & \gamma(0) & 0 & 1 & 0 & \dots & 0 \\
			0 & \dot\gamma(0) & 0 & 0 & 1 & \dots & 0 \\
			\vdots & \vdots & \vdots & \vdots & \vdots & \ddots & \vdots \\
			0 & \gamma^{(n)}(0) & -a_0 & -a_1 & -a_2 &  \dots & -a_n
		\end{pmatrix}.
\end{equation}
The differential equations~\eqref{odes} or~\eqref{odes_vec} do not explicitly contain memory effects. The future of a trajectory is uniquely determined once the present values of the coordinates~$q, p, \zeta_0, \dots, \zeta_n$ have been prescribed. The auxiliary coordinates $\zeta_i$ encode all the relevant information about the history of a trajectory. In order to obtain a closed system of equations for finitely many coordinates, it is essential that the friction kernel satisfies the condition~\eqref{gammaEq}. If no such condition holds, one obtains instead a differential equation system that involves all coordinates~$\zeta_i$ for $0\le i<\infty$. The phase space of the system is then infinite dimensional. Even in this case, however, the construction shows that the phase space dimension is always countably infinite.

The formulation~\eqref{odes} of the GLE in terms of a system of linear differential equations with constant coefficients can be solved with the usual techniques. A solution
\begin{equation*}
	\vec z(t) = \vec z_0\,e^{\epsilon t}
\end{equation*}
with exponential time dependence exists if $\epsilon$ is an eigenvalue of the coefficient matrix~$M$. Martens~\cite{Martens02} has shown that these eigenvalues are exactly the solutions of Eq.~\eqref{evalEq}, so that his construction of the phase space, where it is applicable, leads to the same solutions as the explicit calculation of Sec.~\ref{sec:solInf}.

\section{Solution of the finite-time GLE}
\label{sec:solFin}

The finite-time GLE~\eqref{GLE_finite} poses and initial-value problem: A trajectory is completely determined once the position $q(0)$ and velocity $\dot q(0)$ have been prescribed at time $t=0$. This observation indicates already that the general solution of the finite-time GLE cannot be of the form~\eqref{gensol_inf}, which contains too many free parameters.  Nevertheless, one would expect that the eigenvalues obtained from Eq.~\eqref{evalEq} should still determine the dynamics. Indeed, it will be shown in the following that the phase space of the finite-time GLE can be regarded as a time-dependent surface of the extended phase space that describes the infinite-time GLE. It will also be demonstrated how the extended phase space must be restricted to model the dynamics of the finite-time GLE.

If we once again make an exponential ansatz $q(t) = e^{\epsilon t}$ and substitute into the finite-time GLE, we find that the friction term takes the form
\begin{align*}
	\int_0^t \gamma(t-s) \dot q(s)\,ds
	&= \epsilon \int_0^t \gamma(r) e^{\epsilon(t-r)} \,dr \\
	&= \epsilon e^{\epsilon t} \left(
		\int_0^\infty \gamma(r) e^{-\epsilon r}\,dr - \int_t^\infty \gamma(r) e^{-\epsilon r}\,dr \right) \\
	&= \epsilon e^{\epsilon t} \hat\gamma(\epsilon) 
		- \epsilon \int_0^\infty \gamma(s+t) e^{-\epsilon s}\,ds.
\end{align*}
In comparison to the earlier result~\eqref{damp_inf} for the infinite-time~GLE, there is a correction term that arises from the finite lower limit of integration. Even if $\epsilon$ is chosen to be a solution to the nonlinear eigenvalue equation~\eqref{evalEq}, the exponential ansatz will fail to provide a solution to the GLE, unless the boundary term vanishes for all times~$t$, which is impossible. The finite-time GLE therefore does not have solutions with simple exponential time dependence. 

If more general solutions of the form~\eqref{gensol_inf}
\begin{equation*}
	q(t) = \sum_j c_j e^{\epsilon_j t}
\end{equation*}
are permitted, the condition that the boundary term should vanish reads
\begin{equation}
	\label{finite_cond}
	\sum_j c_j \epsilon_j \int_0^\infty \gamma(t+s) e^{-\epsilon_j s}\,ds = 0.
\end{equation}
This equation, which must be satisfied for all~$t$, provides constraints on the parameters~$c_j$. It therefore reduces the dimension of the phase space below the number (including multiplicities) of eigenvalues~$\epsilon_j$.

The discussion of the previous section allows us to rewrite the boundary condition~\eqref{finite_cond} in a more suggestive manner. Solutions of the finite-time GLE are, by assumption, defined only for times $t\ge 0$. However, a solution of the known analytical form~\eqref{gensol_inf} can easily be extended to negative times. If we take the liberty to do this, \eqref{finite_cond} takes the form
\begin{equation}
	\label{finite_cond2}
	\int_{-\infty}^0 \gamma(t-s) \dot q(s)\,ds = 0,
\end{equation}
which should again hold for all times~$t$. For a trajectory that satisfies this condition, the finite-time and infinite-time friction terms yield the same result. This condition therefore demands that the history of the trajectory before time $t=0$ should have no influence on the friction force at any time $t>0$, which is precisely the assumption by which the finite-time GLE differs from the infinte-time GLE.

By setting $t=0$ in Eq.~\eqref{finite_cond2}, we obtain the condition $\zeta_0(0)=0$, with the auxiliary variable $\zeta_0$ introduced in the previous section. If $\gamma(t)$ is sufficiently differentiable, we can take derivatives of the condition~\eqref{finite_cond2} to obtain
\begin{equation}
	\label{zeta_IC}
	\zeta_i(0)=0\quad \text{for all $i\ge 0$.}
\end{equation}
If the friction kernel satisfies a differential equation of the form~\eqref{gammaEq}, only the first $n+1$ of the conditions~\eqref{zeta_IC} are independent, and the validity of the remaining conditions is enforced by the differential equation.

This observation allows us to interpret the physical origin of the boundary condition~\eqref{finite_cond}: The auxiliary coordinates $\zeta_i$ encode all the information about the past of a trajectory that is necessary to predict its future. A point $(q,p,\zeta_0,\dots,\zeta_n)$ in the extended phase space corresponds to a possible state of the system if and only if a past trajectory can be found for which all variables take the prescribed values. At first glance, it might appear obvious that this should always be possible, at least for the infinite-time GLE, because one has the freedom to adjust the (infinitely many) values of the function $q(s)$ for $s$ from $-\infty$ to the present, and one has to match only a finite or countably infinite number of prescribed values for the phase space coordinates. However, if the function~$q(s)$ is to describe a possible history of the system, it must satisfy the GLE for all past instances of time. This condition drastically reduces the freedom in the choice of a past trajectory, and it is not obvious what precisely the consequences of this requirement will be.

In the case of the infinite-time GLE, the differential equations~\eqref{odes} allow one to calculate the past as well as the future of a trajectory for an arbitrary initial condition $(q,p,\zeta_0,\dots,\zeta_n)$ in the extended phase space. This observation shows that for every such initial condition it is always possible to identify a suitable past trajectory~$q(t)$, and this trajectory is unique. The extended phase space therefore is the dynamical phase space of the system in which every point corresponds to one and only one possible trajectory.

The situation is drastically different in the case of the finite-time GLE. While it is clearly possible to prescribe the values $q(0)$ and $p(0)$ at the initial time $t=0$, one cannot choose arbitrary values for the $\zeta_i$ at this time because the trajectory does not have a past that could be adjusted to the prescribed $\zeta_i$. Indeed, from the modified definition (which is the original definition of Ref.~\cite{Martens02})
\begin{equation}
	\label{zetaDefFin}
	\zeta_i(t) = \int_{0}^t \gamma^{(i)}(t-s) \,p(s) \,ds
\end{equation}
that is appropriate to the finite-time GLE, it is obvious that $\zeta_i(0)$ must always be zero and cannot be chosen arbitrarily. We are thus led back to the condition~\eqref{zeta_IC}. (Note that the difference between the original definition~\eqref{zetaDef} and the modified definition~\eqref{zetaDefFin} is immaterial under the constraint~\eqref{finite_cond2}). As was to be expected, the only initial conditions that can be imposed for the finite-time GLE are the values $q(0)$ and $p(0)$.

\section{The phase space of the finite-time GLE}
\label{sec:finitePhase}

The extended phase space described in Section~\ref{sec:extPhase} cannot be regarded as the true dynamical phase space of the finite-time GLE because its dimension is too high. Instead, the true phase space is a two-dimensional surface within the extended phase space that is characterized by the constraints~\eqref{zeta_IC}. An important subtlety to note about these constraints is that they impose conditions on the  initial values of the auxiliary coordinates $\zeta_i$, not on the current values. To decide if a point in the extended phase space is in the true phase space at time $t$, one has to follow the trajectory through this point back to time $t=0$ and then check the initial values $\zeta_i(0)$. For a given point the initial values of $\zeta_i$  will depend on $t$. As a consequence, the true phase space is a time-dependent surface $\Sigma(t)$ of the extended phase space. This surface, which for the linear GLE~\eqref{GLE_finite} is a plane, will in the following be referred to as the phase plane, as opposed to the extended phase space in which it is embedded. The observation that the finite-time GLE does not permit solutions with purely exponential time dependence is intimately related to this time dependence of the phase plane.

We will now derive equations of motion for the dynamics in the phase plane. It will be shown that these equations are still linear, but explicitly time dependent. Consider the dynamics in the extended phase space that is described by Eq.~\eqref{odes_vec}
\begin{equation*}
	\dot{\vec z} = M \vec z
\end{equation*}
with the matrix~$M$ of Eq.~\eqref{Mdef}. (For the sake of definiteness, we will focus on the case in which the extended phase space has finite dimension. Similar arguments apply to the infinite-dimensional case.) Let $U(t)$ be the time evolution operator of Eq.~\eqref{odes_vec}, i.e., let the dynamics be given by
\begin{equation}
	\label{evol_auto}
	\vec z(t) = U(t)\,\vec z(0).
\end{equation}

Let the operator $P(t)$ project a point $(q,p,\zeta_0,\dots)$ in the extended phase space onto the initial values $(\zeta_0(0), \zeta_1(0),\dots)$ of the auxiliary coordinates. Points in the phase plane $\Sigma(t)$ that represents the dynamical phase space at time $t$ are then described by the condition
\begin{equation}
	\label{projDef}
	\vec z \in \Sigma \quad \Leftrightarrow \quad P(t) \vec z = \vec 0.
\end{equation}
At time $t=0$, this projector has the form
\begin{equation*}
	P(0)=\begin{pmatrix}
			0 & 0 & 1 & 0 & \dots \\
			0 & 0 & 0 & 1 & \dots \\
			\vdots & \vdots & \vdots & \vdots & \ddots
		\end{pmatrix}.
\end{equation*}
Because the $P(t)$ projects onto initial values, it satisfies
\begin{equation*}
	P(0) \vec z(0)  = P(t) \vec z(t) = P(t) U(t) \vec z(0)
\end{equation*}
or
\begin{equation}
	\label{proj_evol}
	P(0) = P(t) U(t) = \text{const}.
\end{equation}
Taking the time derivative of this equation, we find
\begin{equation*}
	 \dot P(t) = - P(t) \dot U(t) U^{-1}(t)  = -P(t) M
\end{equation*}
because $\dot U = M U$.

To introduce a coordinate system in the phase plane $\Sigma$, we choose two orthogonal unit vectors $\vec v_1(t)$ and $\vec v_2(t)$ in $\Sigma$. These vectors have to be time-dependent in such a way that they stay in $\Sigma$ at all times, but this requirement does not determine the time-dependence completely. We can impose the condition that their derivative $\dot{\vec v}_i$ be perpendicular to the plane $\Sigma$, i.e.
\begin{equation}
	\label{vConst}
	\dot{\vec v}_i\cdot\vec v_j = 0 \quad \text{for $i,j=1,2$}.
\end{equation}
This choice also ensures that the orthonormality conditions
\begin{equation}
	\vec v_1^2 = \vec v_2^2 = 1, \qquad \vec v_1\cdot \vec v_2=0
\end{equation}
are satisfied at all times if they are satisfied at $t=0$.

Because the vectors $\vec v_i$ are to be in $\Sigma$, they must satisfy the constraint
\begin{equation*}
	P(t) \vec v_i(t) = 0.
\end{equation*}
Taking the time derivative of this equation, we obtain
\begin{equation*}
	P(t) \dot{\vec v}_i(t) = - \dot P(t) \vec v_i(t) = P(t)\,M\vec v_i(t),
\end{equation*}
whence it follows that
\begin{equation*}
	\dot{\vec v}_i = M \vec v_i + \lambda_{i1} \vec v_1 + \lambda_{i2} \vec v_2
\end{equation*}
with scalar functions $\lambda_{ij}(t)$. The orthogonality conditions~\eqref{vConst} determine these functions as
\begin{equation*}
	\lambda_{ij} = \vec v_j \cdot M\vec v_i,
\end{equation*}
which gives
\begin{equation}
	\label{vEq}
	\dot{\vec v}_i = M \vec v_i - (\vec v_1\cdot M\vec v_i) \vec v_1 
		- (\vec v_2 \cdot M\vec v_i) \vec v_2.
\end{equation}
These equations of motion determine the coordinate vectors $\vec v_i(t)$ completely once $\vec v_1(0)$ and $\vec v_2(0)$ have been arbitrarily chosen as an orthonormal system in $\Sigma(0)$. Note that the right-hand side of Eq.~\eqref{vEq} is simply the component of $M\vec v_i$ that is perpendicular to $\Sigma(t)$. Thus, the coordinate vectors are subject to the dynamics~\eqref{odes_vec}, only corrected to satisfy the constraint~\eqref{vConst}.

Once the coordinate system in the phase plane has been chosen, an arbitrary point in the plane can be described with two coordinates $y_1$ and $y_2$ by
\begin{equation}
	\vec z = y_1 \vec v_1 + y_2 \vec v_2,
\end{equation}
and the dynamics in the phase plane will be described by equations of motion for the $y_i$. These equations can be obtained from~\eqref{odes_vec} by noting that
\begin{align}
	\dot y_i & = \vec v_i \cdot M\vec z \nonumber \\
		&= (\vec v_i\cdot M\vec v_1) y_1 + (\vec v_i\cdot M\vec v_2) y_2
\end{align}
or
\begin{equation}
	\label{yEq}
	\frac{d}{dt} \begin{pmatrix} y_1 \\ y_2 \end{pmatrix} =
		\langle M\rangle_{\Sigma(t)} \begin{pmatrix} y_1 \\ y_2 \end{pmatrix},
\end{equation}
where
\begin{equation}
	\label{MprojDef}
	\langle M\rangle_{\Sigma(t)} =
		\begin{pmatrix}
			(\vec v_1\cdot M\vec v_1) & (\vec v_1\cdot M\vec v_2) \\
			(\vec v_2\cdot M\vec v_1) & (\vec v_2\cdot M\vec v_2)
		\end{pmatrix}
\end{equation}
is the projection of the matrix $M$ that describes the dynamics in the extended phase space onto the instantaneous position of the phase plane $\Sigma$. This projection is explicitly time dependent, and it must be because the plane $\Sigma$ is itself time-dependent. As a consequence, the equations of motion~\eqref{yEq} do not permit solutions with purely exponential time dependence, although they are linear. This confirms the observation made in Section~\ref{sec:solFin}. The development of the present section gives a geometric interpretation of this observation, and it provides an explicit construction of the phase plane that represents the true two-dimensional dynamical phase space of the finite-time GLE.

\section{Example: Exponential memory friction}
\label{sec:exp}

As an example, we will study the case of an exponentially decaying memory kernel
\begin{equation}
	\label{expDamp}
	\gamma(t) = \frac{\gamma}{\tau} e^{-t/\tau}
\end{equation}
with a characteristic memory time $\tau$ and a damping strength $\gamma$. This friction kernel is normalized such that
\begin{equation*}
	\int_{-\infty}^0 \gamma(t) \, dt = \gamma.
\end{equation*}
It satisfies a first-order differential equation of the form~\eqref{gammaEq} and therefore leads to a three-dimensional extended phase space, the smallest dimension possible.

\begin{figure}
	\includegraphics{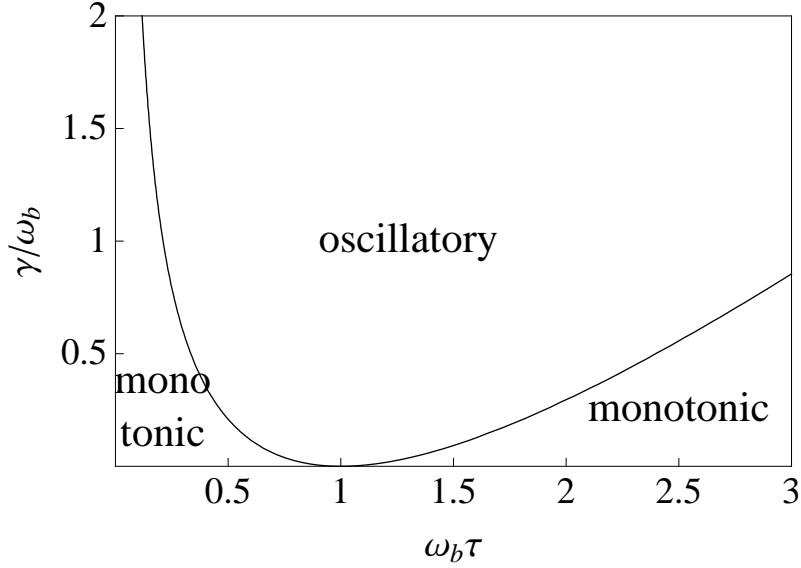}
	\caption{Regions in the parameter space of the infinite-time GLE with exponential damping~\eqref{expDamp} that lead to monotonic decay or damped oscillations in the bath modes.}
	\label{fig:param}
\end{figure}

The geometry of the extended phase space was described in detail in \cite{Martens02}. There is always one positive real eigenvalue. The two remaining eigenvalues have negative real parts and correspond to damped bath modes. They are either both real or form a complex conjugate pair. Excitations in the bath modes either decay monotonically or perform damped oscillations, respectively. The parameter regions that lead to each type of behavior are illustrated in Fig.~\ref{fig:param}.

\begin{figure}
	\includegraphics{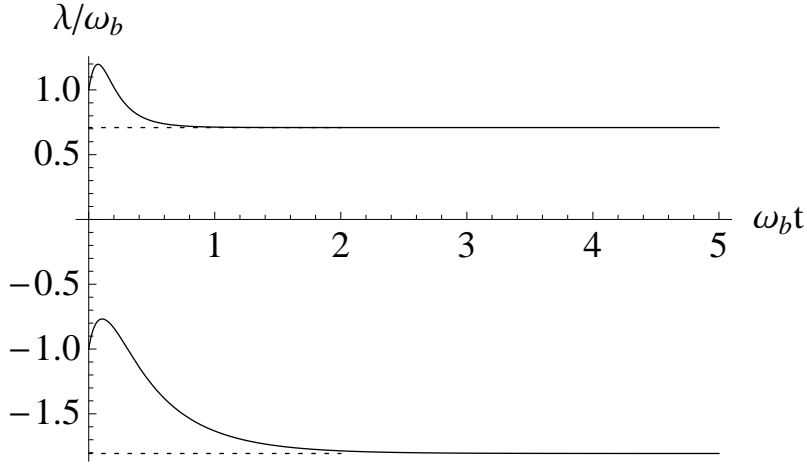}
	\caption{Eigenvalues of the projected dynamics described by $\langle M\rangle_{\Sigma(t)}$ as a function of time for $\gamma=0.8\omb$ and $\omb\tau=0.2$. Dotted lines indicate the two largest eigenvalues of the dynamics in the extended phase space.}
	\label{fig:evalMono}
\end{figure}

The dynamics in the true phase space is obtained by projecting from the extended phase space onto the phase plane $\Sigma(t)$. The motion of this plane is itself determined by the dynamics in the extended phase space. For this reason, in the long time limit the phase plane will, loosely speaking, be aligned with the unstable and the slowest-decaying stable mode. In the case of three real eigenvalues in the extended phase space, this is literally true: For long times, one of the coordinate vectors $\vec v_i$ is parallel to the unstable eigenvector, the second is parallel to the eigenvector with largest negative eigenvalue, subject to the condition that the two vectors must be perpendicular. The influence of  the smallest eigenvalue disappears in this limit. That this is actually the case is shown by the numerical example in Fig.~\ref{fig:evalMono}. The figure shows the eigenvalues of the matrix $\langle M\rangle_{\Sigma(t)}$ that describes the instantaneous dynamics in the phase plane as a function of time. At $t=0$, these eigenvalues are $\pm\omb$. This must be the case for any choice of the friction kernel $\gamma(t)$, as can easily be verified from Eq.~\eqref{Mdef} or from the fact that the finite-time damping term in Eq.~\eqref{GLE_finite} is zero for $t=0$.

The observation that both eigenvalues and eigenvectors of the dynamics within $\Sigma(t)$ converge to fixed limits makes it easy to describe the long-time dynamics. There will be a stable manifold, i.e. a set of trajectories that approach the origin as $t\to\infty$. At sufficiently large $t$, this stable manifold will coincide with the subspace spanned by the stable eigenvector within the phase plane. However, the picture is more complicated at early times, when the time-dependent eigenvectors have not yet reached their limiting values. The stable manifold is itself time-dependent, and at early times it will not be aligned with the instantaneous stable eigenvector. Instead, at these times a full solution of the time-dependent equations of motion is necessary to identify those trajectories that will ultimately approach the origin.

Although there is an unstable eigenvector within the phase plane, an unstable manifold can, strictly speaking, not be defined. Such a manifold is generally defined as the set of trajectories that approach the origin as $t\to-\infty$, but this limit cannot be taken for the finite-time GLE. Nevertheless, it is clear that for $t\to\infty$ all trajectories outside the stable manifold will exhibit exponential instability along the direction of the unstable eigenvector, so that the space spanned by this eigenvector plays the role of an unstable manifold for practical purposes.

\begin{figure}
	\includegraphics{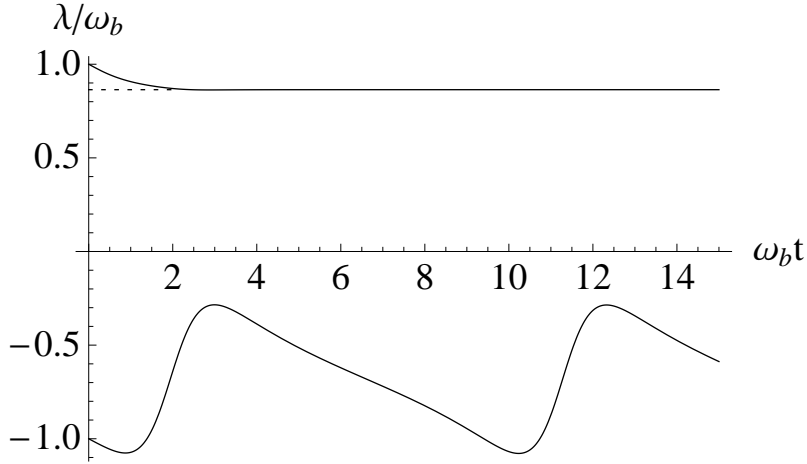}
	\caption{Eigenvalues of the projected dynamics described by $\langle M\rangle_{\Sigma(t)}$ as a function of time for $\gamma=0.8\omb$ and $\omb\tau=2$. Dotted line indicates the unstable eigenvalues of the dynamics in the extended phase space. The other two eigenvalues form a complex conjugate pair.}
	\label{fig:evalOsc}
\end{figure}

A more complex situation, illustrated in Fig.~\ref{fig:evalOsc}, arises if the stable eigenvalues of the extended dynamics form a complex conjugate pair. As before, at time $t=0$ the eigenvalues of the dynamics within the phase plane take the values $\pm\omega_b$. For large time, the unstable eigenvalue approaches the unstable eigenvalue of the extended dynamics and one of the coordinate vectors $\vec v_i$ aligns with the corresponding unstable eigenvector. The second eigenvalue, however, does not approach a fixed limit, but a limit cycle. Because the two stable eigenvalues in the extended phase space describe oscillatory dynamics, the phase plane $\Sigma(t)$ is forced to rotate around the unstable direction. The projected matrix $\langle M\rangle_{\Sigma(t)}$ takes the same value twice in every rotation period because the orientation of $\Sigma(t)$ is irrelevant. Indeed, it can be checked numerically that the oscillation period in Fig.~\ref{fig:evalOsc} is half the period described by the imaginary part of the complex eigenvalues of the extended dynamics.

In the case of three real eigenvalues, it is easy to conclude that for large times the phase plane $\Sigma(t)$ must be spanned by the two eigenvectors that correspond to the leading eigenvalues. No such simple prediction is possible in the case of complex eigenvalues. It is clear that the phase plane will contain the unstable eigenvector and will carry out a rotation in the plane spanned by the two stable eigenvectors. The frequency of that rotation is given by the imaginary part of the eigenvalues, but its phase can only be obtained from an explicit solution of the equations of motion, not from qualitative considerations. The coordinate direction that is aligned with the unstable eigenvector will still play the role of a (quasi-)unstable manifold. The instantaneous stable eigenvector of the dynamics within the phase plane, however, will be periodically time-dependent even in the long-time limit, and it is therefore not obvious where the stable manifold of the origin will be. Again, this can, of course, be determined from a full solution of the equations of motion, but not from a simple qualitative argument.

The exponentially decaying friction kernel that was discussed in the previous section represents the simplest possible case because it leads to the lowest possible dimension of the extended phase space. However, the characteristics of the dynamics within the phase plane that were found in this special case are typical of the general case because any fast decaying additional bath modes that are present in the extended phase space lose their influence on the dynamics in the long-time limit, as was shown in the first scenario of Sec.~\ref{sec:exp}. 

The motion of the phase plane, which is represented by the motion of the coordinate vectors $\vec v_i$, is determined by the dynamics in the extended phase space. In the long-time limit, one of the coordinate vectors will always be aligned with the unstable eigenvector. The second coordinate vector will be determined by the eigenvector or eigenvectors corresponding to the bath modes that decay most slowly. If there is a single such eigenvector with a negative real eigenvalue, the coordinate vector will be aligned with it (as far as possible subject to the constraint that it must be perpendicular to the first coordinate vector), and the dynamics in the phase plane will qualitatively show the behavior of the first scenario in Sec.~\ref{sec:exp}. If the slowest-decaying mode is a damped oscillation, the phase plane will rotate around the unstable eigenvector as described by this mode, and the dynamics within the phase plane will follow the pattern of the second scenario in Sec.~\ref{sec:exp}.

The two cases discussed above will therefore correctly describe the dynamics in the phase plane of the finite-time GLE in the long-time limit for arbitrary friction kernel, except possibly in degenerate cases where several decaying or oscillatory bath modes have eigenvalues with the same real part. For short times, of course, the influence of the strongly damped modes is not negligible. If one's goal is, for example, to find stable manifolds at time $t=0$, it is necessary to take full account of all phase space dimensions, and the simplifications that are afforded by the long-time limit are not available.

\section{Inclusion of the noise}
\label{sec:noise}

The most important feature of the generalized Langevin equation~\eqref{GLE_finite_inh} and~\eqref{GLE_inf_inh} is the noise term. It might appear odd, therefore, that throughout this work the noise term has been neglected. It is crucial to see how the inclusion of the noise will change the picture that has so far been presented.

It is straightforward to incorporate the noise into the phase space formulation of the infinite-time GLE. A noise term will appear in the second equation of the system~\eqref{odes}, which will then read
\begin{equation*}
	\dot p = \omb^2 q - \zeta_0 + \xi_\alpha(t).
\end{equation*}
The equations of motion~\eqref{odes_vec} are replaced by
\begin{equation}
	\label{odes_noisy}
	\dot{\vec z} = M \vec z + \vec\xi_\alpha(t),
\end{equation}
where
\begin{equation*}
	\vec\xi_\alpha(t) = \begin{pmatrix} 0 \\ \xi_\alpha(t) \\ 0 \\ \vdots \\ 0 \end{pmatrix}.
\end{equation*}
The noise changes individual trajectories, but the dynamics still take place in the extended phase space spanned by $q, p, \zeta_0, \dots, \zeta_n$, or possibly $q,p$ and infinitely many $\zeta_i$.

The case of the finite-time GLE is, once again, not as simple.  A trajectory is determined by two initial conditions and  the realization of the noise that is acting upon it. The phase space of the finite-time GLE must therefore still be two-dimensional, but if it is regarded as embedded into the extended phase space, its embedding will depend on the noise. 

Specifically, the autonomous time evolution~\eqref{evol_auto} in the extended phase space is replaced by a time evolution
\begin{equation}
	\label{evol_driven}
	\vec z(t) = U_\alpha (t; \vec z(0))
\end{equation}
that depends explicitly on the realization~$\alpha$ of the noise. The time evolution is no longer linear, and therefore cannot be represented by a matrix, because the noisy equations of motion are inhomogeneous. As in the noiseless situation, the phase space of the finite-time GLE at time $t=0$ is the plane $\Sigma(0)$ that contains all points in the extended phase space that satisfy $\zeta_i=0$. At later times, the phase space contains all points that evolve out these initial conditions, i.e. the phase space of the finite-time GLE is obtained through the time evolution of $\Sigma(0)$ as
\begin{equation}
	\label{noisyPlane}
	\Sigma_\alpha(t) = U_\alpha(t; \Sigma(0)).
\end{equation}
Significantly, the phase plane that is obtained in this way is not only time-dependent, as it was in the noiseless version of the finite-time GLE, it also depends on the realization~$\alpha$ of the noise.

The nonlinear time evolution~\eqref{evol_driven} for the noisy dynamics on a harmonic barrier can be obtained explicitly by the same approach that was used in \cite{Bartsch05b,Bartsch05c} to obtain a geometric description of Transition State Theory for the Langevin equation: The solution of the noisy equation of motion~\eqref{odes_noisy} is split into two components
\begin{equation}
	\label{TS_split}
	\vec z(t) = \vec z_\alpha^{\ddag}(t) + \Delta\vec z(t).
\end{equation}
Here $\vec z_\alpha^{\ddag}(t)$ is a particular solution of the noisy equation~\eqref{odes_noisy} that remains in the vicinity of the barrier top at all times and is called the Transition State trajectory. As the subscript~$\alpha$ indicates, it depends on the noise. Explicit formulas for $\vec z_\alpha^\ddag(t)$ as an integral over the noise are given in~\cite{Bartsch05b,Bartsch05c} (where the formulas require knowledge of the noise at times $t<0$, it can be chosen arbitrarily for the finite-time case). The relative coordinate $\Delta\vec z(t)$ satisfies the noiseless equation of motion
\begin{equation*}
	\Delta\dot{\vec z} = M \,\Delta\vec z
\end{equation*}
and therefore follows the noiseless time evolution
\begin{equation*}
	\Delta \vec z(t) = U(t) \, \Delta\vec z(0)
\end{equation*}
with the same linear time evolution operator as in Eq.~\eqref{evol_auto}. We can then write
\begin{align*}
	\vec z(t) &= U(t) \, \Delta\vec z(0) + \vec z_\alpha^\ddag(t) \\
		&= U(t) \, (\vec z(0) - \vec z_\alpha^\ddag(0)) + \vec z_\alpha^\ddag(t) \\
		&\equiv U_\alpha(t; \vec z(0)) ,
\end{align*}
which gives an explicit expression for the nonlinear time-evolution operator defined by Eq.~\eqref{evol_driven}. It can be solved for
\begin{equation*}
	\vec z(0) = U^{-1}(t) \vec z(t) - U^{-1}(t) \vec z_\alpha^\ddag(t) + \vec z_\alpha^\ddag(0).
\end{equation*}
The condition for a point $\vec z(t)$ to be in the phase plane of the finite-time GLE at time $t$ can then be stated as
\begin{align*}
	\vec 0 &= P(0)\,\vec z(0)  \nonumber \\
		&= P(t)\,\vec z(t) - P(t)\,\vec z_\alpha^\ddag(t) + P(0)\,\vec z_\alpha^\ddag(0), \nonumber \\
		&\equiv P(t)\,\vec z(t) - \vec s_\alpha(t),
\end{align*}
where the time evolution Eq.~\eqref{proj_evol} of the projector $P(t)$ has been used. This condition, which characterizes the phase plane $\Sigma_\alpha(t)$, can be restated as
\begin{equation}
	 P(t)\,\vec z(t) = \vec s_\alpha(t)
\end{equation}
with the vector
\begin{equation}
	\label{sDef}
	\vec s_\alpha(t) =  P(t)\,\vec z_\alpha^\ddag(t) - P(0)\,\vec z_\alpha^\ddag(0).
\end{equation}
It describes a time-dependent plane that is parallel to the phase plane $\Sigma(t)$ for the noiseless case that was defined in~\eqref{projDef}, but it is shifted away from the origin by an amount that is stochastically time dependent. As anticipated, therefore, the phase space of the finite-time GLE for any given realization of the noise is a two-dimensional plane within the extended phase space.

An arbitrary vector within the phase plane can be written as
\begin{equation}
	\vec z = y_1 \vec v_1 + y_2 \vec v_2 + \vec S_\alpha
\end{equation}
where the two vectors $\vec v_1(t)$ and $\vec v_2(t)$ span the noiseless phase plane $\Sigma(t)$ (see Section~\ref{sec:finitePhase}) and the reference vector $\vec S_\alpha(t)$ satisfies $P(t) \vec S_\alpha(t) = \vec s_\alpha(t)$. Eq.~\eqref{odes_noisy} then translates into the equations of motion
\begin{equation}
	\dot y_i = \vec v_i \cdot (M\vec v_1) y_1 + \vec v_i \cdot (M\vec v_2) y_2 
			+ \vec v_i\cdot \vec \xi_\alpha(t) - \vec v_i \cdot \dot{\vec S}_\alpha(t)
\end{equation}
or
\begin{equation}
	\label{yEqNoisy}
	\frac{d}{dt} \begin{pmatrix} y_1 \\ y_2 \end{pmatrix} =
		\langle M\rangle_{\Sigma(t)} \begin{pmatrix} y_1 \\ y_2 \end{pmatrix} 
		+ \langle \vec\xi_\alpha(t)\rangle_{\Sigma(t)} 
		- \langle \dot{\vec S}_\alpha(t)\rangle_{\Sigma(t)},
\end{equation}
where $\langle M\rangle_{\Sigma(t)}$ is, as in~\eqref{MprojDef}, the projection of the matrix $M$ onto the plane $\Sigma(t)$ and the last two terms denote the projections of the given vectors onto that plane. In comparison to the noiseless equation of motion~\eqref{yEq}, the term $\langle \vec\xi_\alpha(t)\rangle_{\Sigma(t)}$ describes precisely the impact of the noise on the dynamics that one would expect. The unexpected term $- \langle \dot{\vec S}_\alpha(t)\rangle_{\Sigma(t)}$ arises from a motion of the reference vector $\vec S_\alpha(t)$ along the instantaneous direction of the phase plane. This term can be eliminated by a suitable redefinition of $\vec S_\alpha(t)$: If an arbitrary vector in $\Sigma(t)$ is added to $\vec S_\alpha(t)$,  the shifted phase plane $\Sigma_\alpha(t)$ remains unchanged.

The formalism presented in this section can directly be generalized to systems with anharmonic barriers. In this case, the finite-time GLE~\eqref{GLE_finite_inh} will lift to a nonlinear equation of motion
\begin{equation}
	\label{ode_general}
	\dot{\vec z} = \vec f(\vec z)
\end{equation}
in the extended phase space. As before, this equation of motion must be solved subject to the constraint $\zeta_i(0)=0$ on the auxiliary variables, and this constraint defines a two-dimensional plain $\Sigma(0)$ that represents the phase space at time $t=0$. The solution of the equation of motion~\eqref{ode_general} is described by a nonlinear and stochastic time evolution operator $\vec z(t) = U_\alpha(t; \vec z(0))$ as in Eq.~\eqref{evol_driven}. The phase space $\Sigma_\alpha(t)$ at a time $t>0$ is given by the time evolution of the initial phase plane $\Sigma(0)$ according to Eq.~\eqref{noisyPlane}
\begin{equation*}
	\Sigma_\alpha(t) = U_\alpha(t; \Sigma(0)).
\end{equation*}
Because the time evolution operator is nonlinear, the phase space $\Sigma_\alpha(t)$ will in general not be a plane. For an arbitrary anharmonic potential in the GLE, it will usually be impossible to obtain explicit formulas for either the time evolution operator $U_\alpha$ or  the phase space $\Sigma_\alpha(t)$. Nevertheless, in qualitative terms the situation remains unchanged: The dynamical phase space of the finite-time GLE is represented by a two-dimensional submanifold of the extended phase space that is moving stochastically through the larger extended phase space.

\section{Concluding remarks}
\label{sec:conc}

The development presented here demonstrates that there are important differences between the dynamics of the finite-time and infinite-time GLE. While the infinite-time GLE allows for trajectories with purely exponential time dependence, as one would expect for a linear equation of motion, the finite-time GLE imposes boundary conditions that prohibit such behavior. These conditions reduce the phase space dimension for the finite-time GLE to two. From a geometric point of view, the dynamics of the finite-time and infinite-time GLE differ drastically. The infinite-time GLE can be rewritten as a set of time-independent ordinary differential equations in a phase space with dimension between three and infinity. The finite-time GLE, by contrast, has a two-dimensional phase space, the dynamics in which is described by differential equations that are time-dependent in a complicated way. 

In spite of these dramatic differences in the geometric framework, the dynamics of the finite-time and infinite-time GLE become similar when one studies individual trajectories in the long-time limit. In both cases, the behavior of a trajectory in this limit is determined by one unstable eigenvalue (which is the same in the extended and two-dimensional phase spaces). As the boundary condition imposed by the finite-time GLE recedes into the past, its influence on a particular trajectory becomes less and less palpable, although the differences between the respective phase spaces, which represent the totality of possible trajectories, remain for arbitrarily long times.

%\bibliography{Bibliography}

\end{document}